\documentclass[10pt,journal,compsoc]{IEEEtran}
\usepackage[utf8]{inputenc}
\usepackage{tikz}
\usepackage{amsmath}
\usepackage{mdwlist}
\usepackage{comment}
\usepackage{subfigure}
\usepackage{color}
\usepackage{pifont}
\usepackage{algorithm}
\usepackage{algorithmic}

\usepackage{tikz}
\usepackage{xcolor}
\newcommand*\circled[1]{\tikz[baseline=(char.base)]{
            \node[shape=circle,fill,inner sep=1pt] (char) {\textcolor{white}{#1}};}}

\def\smallerspacecaption{\vspace{-1mm}}

\ifCLASSOPTIONcompsoc
  \usepackage[nocompress]{cite}
\else
  \usepackage{cite}
\fi

\begin{document}
\title{Hardware Trojan Threats to Cache Coherence in Modern 2.5D Chiplet Systems}

\author{Gino A. Chacon,
        Charles Williams,
        Johann Knechtel,
        Ozgur Sinanoglu,
        and Paul V. Gratz
\thanks{G. A. Chacon, C. Williams, and P. V. Gratz are with Texas A\&M University (e-mail: ginochacon@tamu.edu, charlesw2000@tamu.edu, pgratz@gratz1.com).}\protect \thanks{J. Knechtel and O. Sinanoglu are with New York University Abu Dhabi (e-mail: johann@nyu.edu, ozgursin@nyu.edu).}
}%

\IEEEtitleabstractindextext{
\begin{abstract}
  As industry moves toward chiplet-based designs, the insertion of
  hardware Trojans poses a significant threat to the security of these
  systems.  These systems rely heavily on cache coherence for coherent
  data communication, making coherence an attractive target.
  Critically, unlike prior work, which focuses only on malicious
  packet modifications, a Trojan attack that exploits coherence can
  modify data in memory that was never touched and is not owned by the
  chiplet which contains the Trojan.  Further, the Trojan need not
  even be physically between the victim and the memory controller to
  attack the victim's memory transactions.  Here, we explore the
  fundamental attack vectors possible in chiplet-based systems and
  provide an example Trojan implementation capable of directly
  modifying victim data in memory. This work aims to highlight the
  need for developing mechanisms that can protect and secure the
  coherence scheme from these forms of attacks.

\end{abstract}

}



\maketitle

\IEEEdisplaynontitleabstractindextext

\IEEEpeerreviewmaketitle





\IEEEraisesectionheading{\section{Introduction}\label{sec:introduction}}

\IEEEPARstart{C}{omputing} systems are moving toward 2.5D designs that
source various hard IPs, called chiplets, from multiple vendors and
integrate them using an interposer. Industry has demonstrated that
2.5D designs lower manufacturing costs, enabling further scaling post-Moore's
Law~\cite{naffziger21}. Future 2.5D designs may leverage
standards such as Compute Express Link~\cite{cxl} to interoperate
via a shared memory system. While 2.5D designs provide many
benefits, we show they also increase the risk of Trojan attacks,
specifically targeting the coherence system. Here, we
demonstrate several novel Trojans attacking cache coherence in
2.5D designs.  We illustrate the risks for these systems and hope to
excite the architecture community to address these risks. Though
we focus on 2.5D integrated systems, note that these attacks also
apply to general cache-coherent systems integrating closed-source or
hard IP blocks from various vendors.


\emph{Hardware Trojans}, or Trojans for
short~\cite{bhunia2018hardware}, are a threat in which an attacker
infiltrates some level of the design or fabrication process to insert
malicious circuitry into a design.  Trojans can cause disastrous
system failures via confidentiality, integrity, and/or availability
violations. Prior work has shown that Trojans can leak data from
memory~\cite{9061138}, disrupt cryptographic security
features~\cite{7753274}, and induce denial-of-service
attacks~\cite{7926975}.  As industry moves towards 2.5D designs
integrating multiple vendor chiplets, specific chiplets used in
building these systems may be untrustworthy.  Even if the IP vendor is
trustworthy, the manufacturing process may not be, leading to
infiltration and the insertion of Trojans.

In 2.5D designs, memory coherence is crucial to allow each component and 
chiplet to maintain an up-to-date view of the system's memory. We identify this 
system as an ideal target for Trojan attacks as coherence mechanisms control 
how all components communicate data updates. Existing coherence schemes do not 
enforce existing virtual/physical memory permissions, thus, a Trojan connected 
to the coherence scheme can directly manipulate any memory region in the full 
system regardless of memory permissions or physical location. Unlike prior 
packet-level NoC attacks, Trojans on cache coherence do not need to be 
physically on the path between the victim and the memory controller to launch 
effective attacks. Despite this attractiveness, there is a lack of works deeply 
exploring coherence exploits and their defense in 2.5D systems or otherwise.




Here, we propose several new Trojan attacks that leverage the
coherence system protocol to maliciously manipulate the victim
process' memory.  We first describe a set of new fundamental attacks
that a Trojan can mount on coherence systems, \emph{passive reading},
\emph{masquerading}, \emph{modifying}, and \emph{diverting} attacks,
according to Basak \emph{et al.}'s taxonomy~\cite{basak2017ifs}.  Here
we assume an attacker implements these coherence system attacks in
hardware through compromised design or manufacturing.  While each of
these attacks individually violates a system's security, we further
show that adversaries can orchestrate them together to perform complex
\emph{Forging} attacks that modify \emph{any} process' memory.  These
purely hardware attacks \textit{cannot} be thwarted by contemporary
software defense mechanisms since all exploited coherence interactions
are transparent to software and legal within the coherence protocol.
Further, no prior work considers such attacks on coherence systems,
neither in the context of 2.5D systems with chiplets nor traditional
2D systems.

\emph{Contributions.} This work provides new insights into how Trojans
can manipulate coherence systems to violate the security of a chiplet
system.  We present a simulated example of a substantial attack that
can directly manipulate memory in an address space other than that of
the compromised chiplet. This work makes the following contributions:
\begin{itemize}
\item We present potential attack stages available to a 
    Trojan designers exploiting coherence systems.
\item We demonstrate how to use these fundamental stages to
  orchestrate a complex Trojan attack in a chiplet-based
  system.
\item We provide suggestions for future work on 
  hardening modern chiplet designs.
\end{itemize}



\section{Background}
\label{BackgroundMotivation}

\subsection{Hardware Trojans}
\label{hardwaretrojans}

Hardware Trojans are malicious hardware inserted by an attacker during
a device's design or manufacturing process. Chiplet-based devices have
a complex multistage design flow that can be compromised at many
levels, especially as multiple 3rd party vendors emerge to provide
chiplets from separate foundries and design teams. This design flow
makes verification much more difficult and expensive than
traditional System-on-Chip (SoC) manufacturing.

Detecting Trojans is challenging as chiplet-based systems contain
multiple complex IPs sourced from various vendors. Testing a chiplet's
functionality may occur during the manufacturing or integration stage,
which requires a reference model or
device~\cite{bhunia2018hardware}. However, if the 3rd
party IP's source is untrustworthy, the reference model itself may incorporate 
the Trojan, or the IP may camouflage the Trojan as a correct implementation. 
Attackers can conceal a Trojan by only allowing it to trigger under specific 
conditions. For example, the Trojan we describe in Sec.~\ref{forging} only 
activates when it observes references to a specific address. These properties 
make the Trojan difficult to detect by simply testing the functionality of the
chiplet. For this work, we assume that an attacker infiltrates some
stage within the design or manufacturing process to target the
system's coherence mechanisms.


Prior art focuses on infiltrating the NoC of a target design to cause
deadlocks~\cite{7926975}, leak information~\cite{9061138}, or disrupt
security features~\cite{7753274}.  However, NoC-based attacks require
the Trojan to directly attack NoC packets, limiting the Trojan to only
packets traversing a particular path in the NoC.  Prior attacks would
not work in a 2.5D integrated system because attacking chiplets are
not on the path between victim chiplets and the memory controller.


\subsection{Coherence Protocols}

Multi-processor systems incorporate cache coherence protocols to
ensure data coherency across processors' private caches. The coherence
system has a complete vantage point and control over the memory
system. All communication between cores and main memory follows the
coherence protocol, making it an ideal target for a Trojan co-located
with a processor's private caches. At this location, a Trojan can
snoop on coherence messages produced by other processors, manipulate
those messages, or generate messages without incurring exceptions and
remaining invisible to software running on the system.  Further,
coherence schemes do not enforce virtual/physical memory permissions,
thus any Trojan connected to the coherence scheme can directly
manipulate any memory region in the full system, without regard to
memory permissions or physical location.




We target the \emph{MOESI Hammer} coherence protocol, a hybrid
broadcast-directory system used in many AMD
processors~\cite{Conway_2010}.  
Though our focus is MOESI Hammer, our attack scenarios can easily be
ported to other coherence schemes.  In schemes without broadcast, the
Trojan simply needs to register ``S'' or ``X'' state with the
directory to ensure that update requests on the given line are
seen.

\section{Trojans Targeting Coherence Systems}

Here we propose a new set of ``basic'' Trojan attacks on the coherence
scheme that follows the general, not coherence
specific, taxonomy for Trojan attacks by Basak \emph{et
  al.}~\cite{basak2017ifs}: \emph{passive reading},
\emph{masquerading}, \emph{modifying}, and \emph{diverting}. These
basic attacks can adversely affect the system but cannot provide an
attacker with control the memory system.
We then propose a novel, more sophisticated and powerful `\emph{`Forging''}
attack to modify data belonging to another core, even on a different
chiplet than that holding the Trojan.

\noindent \textbf{Target system:} We demonstrate 
our attacks on a 64-core processor with eight chiplets, eight cores per 
chiplet, based on the Rocket-64 architecture~\cite{rocketchip}. Each core has 
private L1 and L2 caches with a unified cache controller. An NoC connects each 
chiplet and four memory controllers that maintain a portion of the global state 
directory.  The cache controllers generate coherence messages that are injected 
into the NoC as network packets.

\subsection{Basic Trojan Attacks on the Coherence Systems}

\begin{figure}[ht]
  \centering
  \subfigure[\textbf{Passive Reading:} Trojan passively observes write
      traffic for other chiplets.  \textbf{(1)} Misses from Chiplet A
      cause \textbf{(2)} broadcast invalidations to all chiplets;
      \textbf{(3)} Trojan snoops invalidation addresses.]
      {\label{fig:passive_reading_attack}\includegraphics[width=0.48\columnwidth]
      {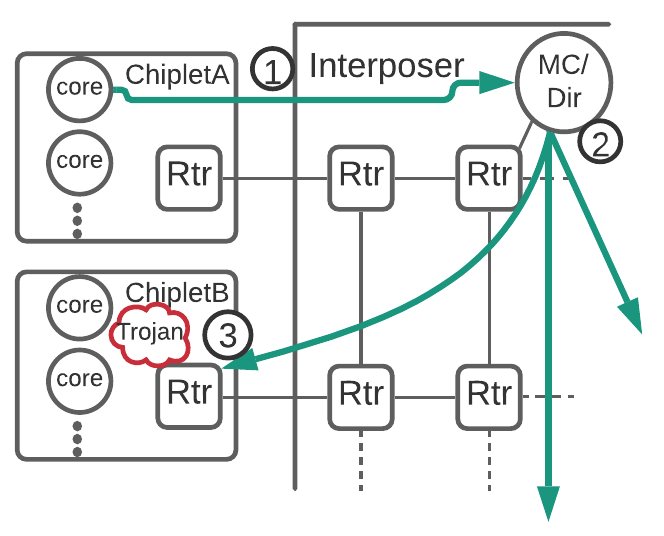}}
  \hfill
  \subfigure[\textbf{Masquerading:} Trojan acts as another
    core. \textbf{(1)} Miss causes GETX to directory; \textbf{(2)}
    broadcast invalidations to each chiplet; \textbf{(3)} Trojan
    blocks local observation, replies with different core ID;
    \textbf{(4)} requesting core proceeds, leaving local caches
    incoherent.]{\label{fig:masquerade_attack}\includegraphics[width=0.48\columnwidth]{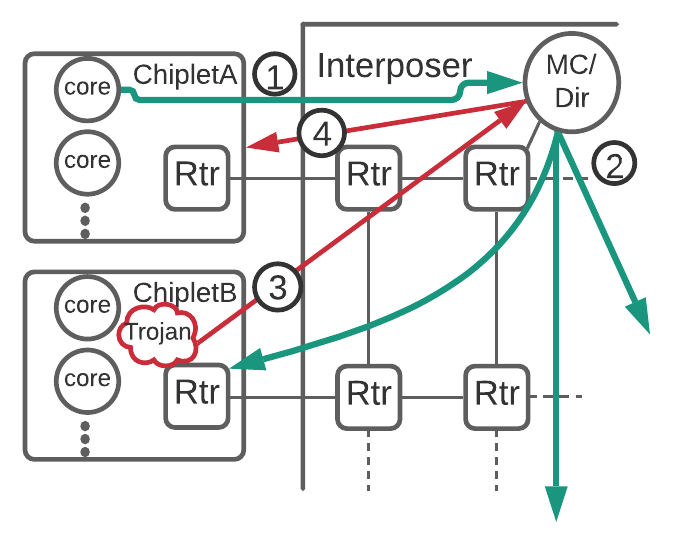}}
  \subfigure[\textbf{Modifying:} Trojan modifies a message to achieve
    incoherent state. \textbf{(1)} Chiplet A sends GETS to
    directory; \textbf{(2)} directory forwards request to Trojan's
    core which has line in `E' state.  Trojan blocks GETS and
    \textbf{(3)} replies with GETX to requestor, \textbf{(4)}
    invalidating Chiplet A's cache entry, leaving attacker in
    control of another cache's contents.]{\label{fig:modifying_attack}\includegraphics[width=0.48\columnwidth]{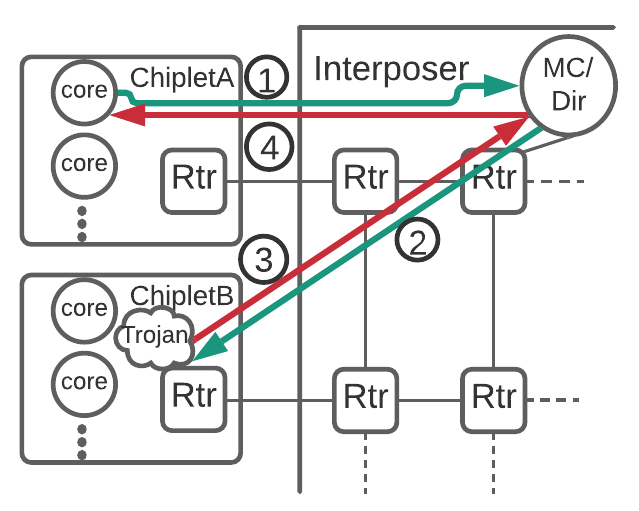}}
  \hfill
    \subfigure[\textbf{Diverting:} Trojan diverts invalidation
      requests. \textbf{(1)} Chiplet A sends GETX to the directory;
      \textbf{(2)} directory broadcasts invalidations. \textbf{(3)}
      Trojan blocks message and diverts a request to another core,
      \textbf{(4)} which responds with a negative-acknowledge or
      acknowledgment resulting in \textbf{(5)} the directory
      allowing original requestor to continue.]{\label{fig:divert_attack}\includegraphics[width=0.48\columnwidth]{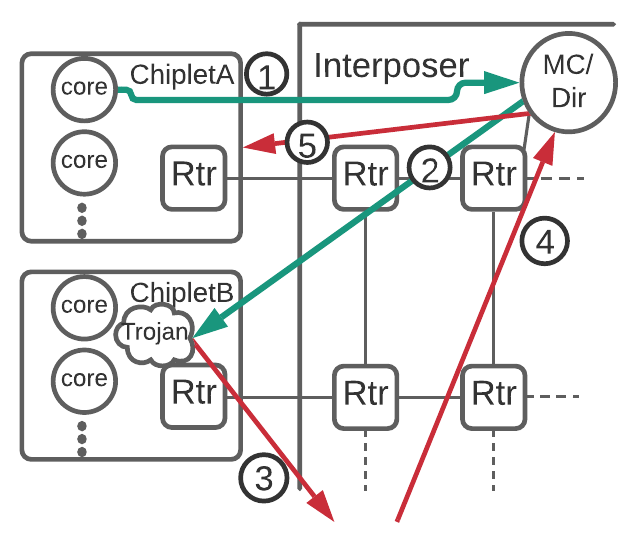}}
  \caption{Coherence Trojan attacks in interposer-based systems in which Chiplet A is the victim of a Trojan attack from Chiplet B.
  }
  \smallerspacecaption
  \smallerspacecaption
\label{fig:coherence_attacks}
\end{figure}

Figure~\ref{fig:coherence_attacks} illustrates our proposed basic
coherence attacks. We assume a Trojan is placed at a core's cache
controller and can intercept coherence messages from the network
interface ahead of the state directory. While these attacks target the
the MOESI Hammer hybrid protocol, the basic principles of the attacks,
are consistent with any coherence protocol.

\noindent\textbf{Passive Reading
  (Fig.~\ref{fig:passive_reading_attack}):}
Trojans passively reading (\textit{snooping}) observe incoming
coherence messages from the chiplet's network-on-chip (NoC)
sub-system as they reach the L2's state directory. The Trojan may
buffer messages, identify specific request patterns, and facilitate a
covert communication channel. The Trojan does not affect the system's
state but may trigger a more complex Trojan. 

\noindent\textbf{Masquerading
	(Fig.~\ref{fig:masquerade_attack}):}
Masquerading (\textit{spoofing}) occurs when a Trojan modifies the
packet's \texttt{sender} field such that the packet appears to
originate from a different core. If the target packet is a request,
such an attack can result in a deadlock since all responses from the
directory or other cores are sent to the incorrect core.  If the
target packet is a response, the Trojan may block it and respond with
an acknowledgment that appears to be from a different core, resulting
in an incoherent memory state.

\noindent\textbf{Modification
	(Fig.~\ref{fig:modifying_attack}):}
Such attacks occur when the Trojan directly modifies the
\texttt{message type} of a coherence message. This attack may result
in a deadlock since the Trojan may cause the memory controller's
directory to assume the data is in one state, due to a modified
packet, while the local directory holds the data in a
different---incorrect---state.

\noindent\textbf{Diverting
	(Fig.~\ref{fig:divert_attack}):}
Trojans can launch diverting attacks by blocking the local state
directory from observing a request and then resending the request with
a different \texttt{destination} field.  This results in the
compromised core and the original requestor becoming incoherent with
respect to the rest of the memory system.

\noindent \textbf{Limitations of Basic Attacks:} Any of the above
attacks can individually result in incoherence or deadlocks but cannot
directly manipulate a victim's data. Only combining these attacks
allows for a more complex set of attack vectors that would enable a
Trojan to pose a significant security threat.







\subsection{Trojan Design}
\label{forging}

\begin{figure}
  \centering \subfigure
  {\label{fig:forging_phase1}\includegraphics[width=0.35\textwidth]{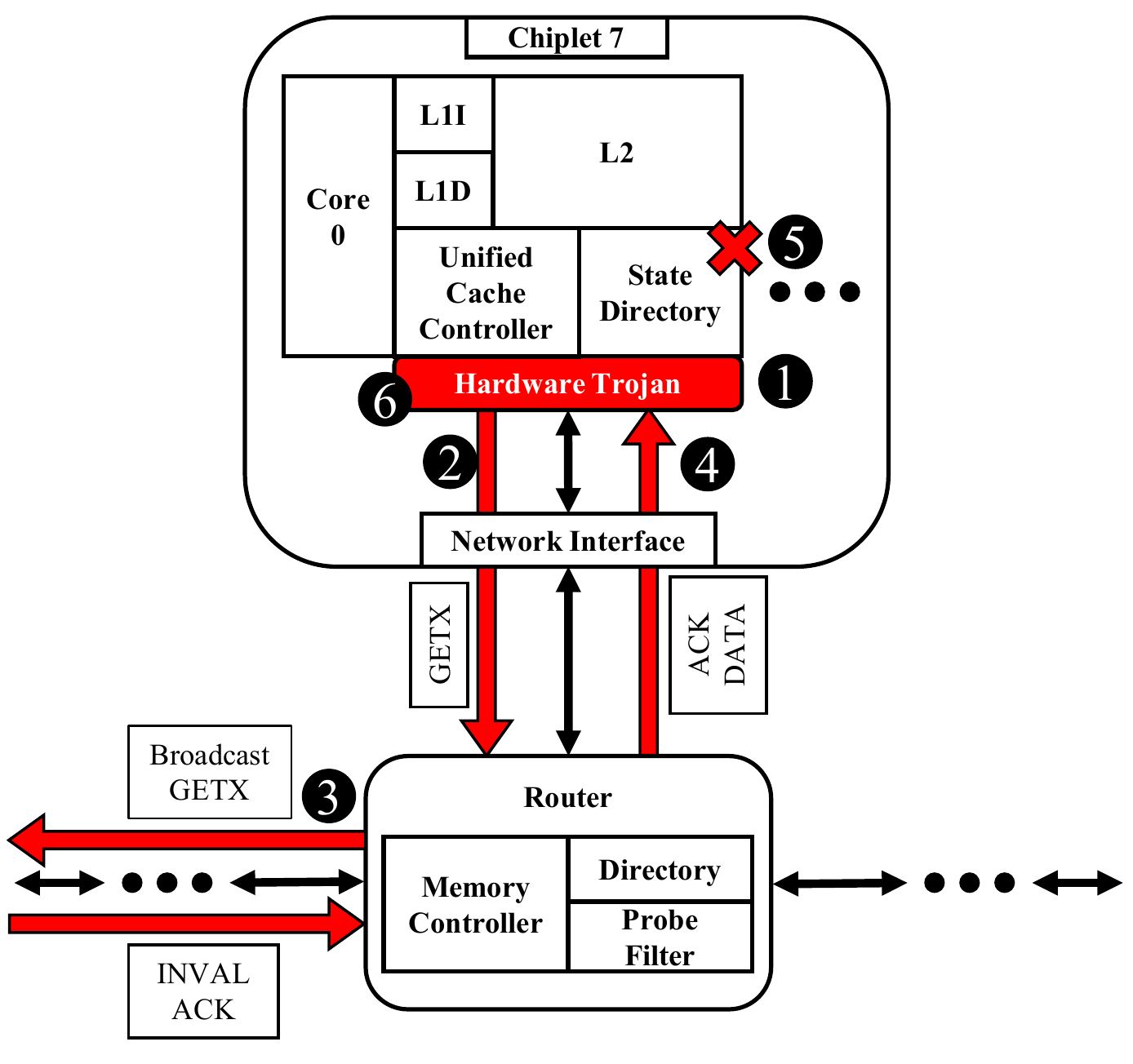}}

  \subfigure
  {\label{fig:forging_phase2}
    \includegraphics[width=0.35\textwidth]{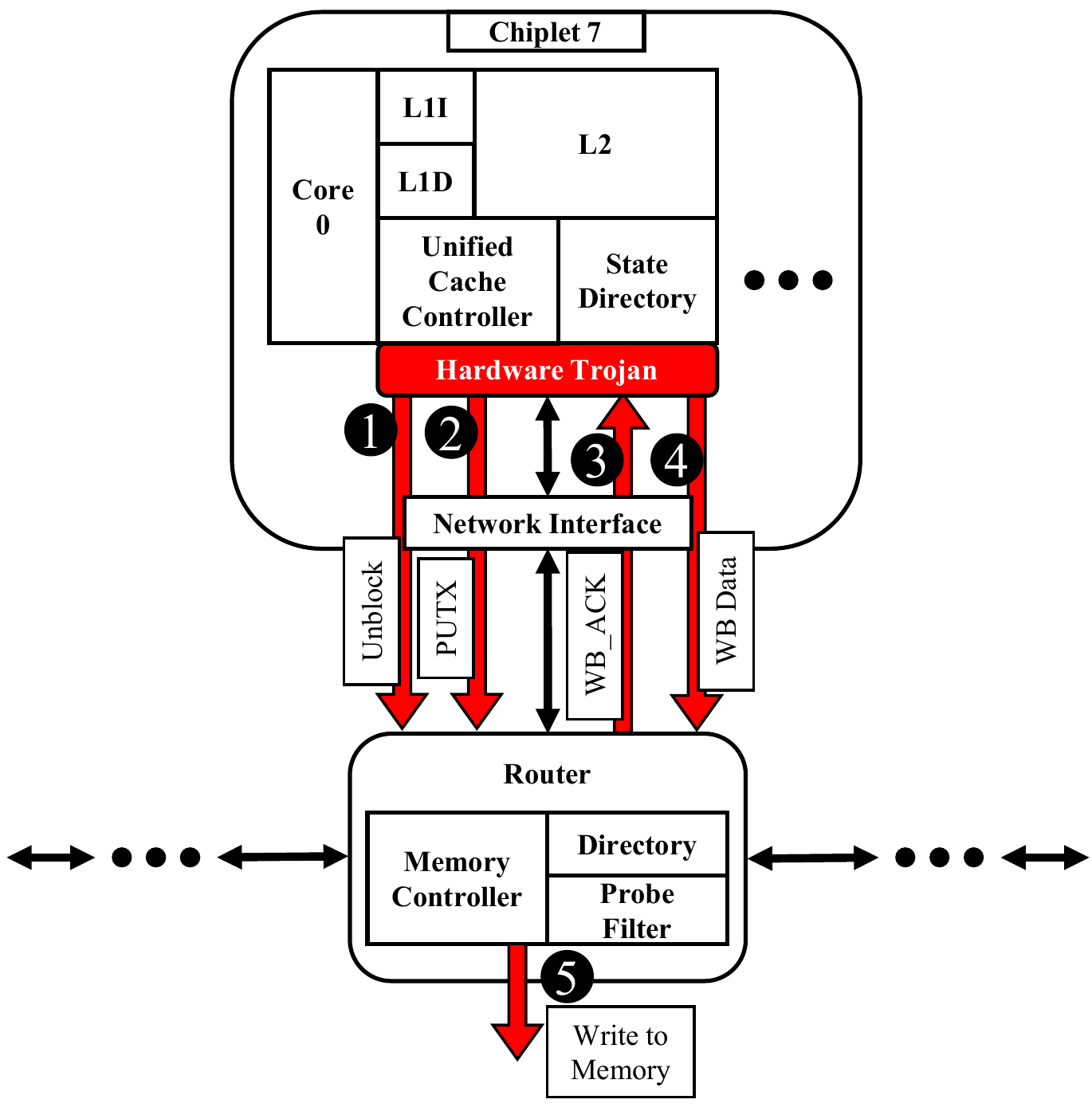}}
  \caption{Our proposed Trojan attack on the coherence system that
    forges messages to gain control and modify specific addresses
    accessed by a process operating in a different chiplet. The attack
    executes in two phases. The first phase (top) allows the Trojan to
    gain control of the target address and the second phase (bottom)
    enables it to mimic the steps required to write back maliciously
    formed data to main memory.}
\label{fig:forging_attack}
\end{figure}

In the remainder of this paper, we propose the \emph{Forging Attack}, a novel attack that manipulates 
legal coherence transactions to allow a Trojan to write to a target address 
in a different process operating in a different chiplet. The compromised 
chiplet containing the Trojan does not have access to the victim process' 
address space but can observe coherence interactions broadcasted by the MOESI 
Hammer protocol. Since the Trojan resides between the network interface and a 
core's state directory in the compromised chiplet (Figure~\ref{fig:forging_attack}),
the Trojan has a complete view of 
incoming or outgoing coherence messages, enabling it to block the core from 
observing specific interactions.
The Trojan holds few registers to track the 
target data's current state relative to the Trojan. These registers imitate the 
core's state directory to ensure the Trojan correctly responds to the global 
directory. 

%
%

\subsection{Forging Attack Demonstration}


Here we assume the Trojan has a predefined target address. In a
real-world scenario, the Trojan can observe coherence messages
broadcasted to the compromised chiplet of the network to select its
target. The coherence protocol requires that the global directory
sends invalidation messages each time a core sends a write request, or
GETX, to a line that it does not own. The invalidation broadcast
removes all copies in other cores before updating the line with new
data.

The Trojan operates in two phases. During the first phase,
the Trojan deceives the global directory into giving the Trojan access
to the data. During the second phase, the Trojan follows the
protocol's required transactions to write to the target address, which
the victim will later read. The interactions caused by the Trojan in
both phases are legal from the perspective of the global
directory. Furthermore, they are transparent to the software executing in
the victim process and all other security software in the system.

\textbf{Phase 1, Acquiring Access to Target Data:} 
Figure~\ref{fig:forging_attack}(top) illustrates the initial steps the Trojan 
must take to gain access permissions to the target address before it can 
maliciously write to it.
\circled{1}~The Trojan observes coherence requests, waiting for a specific 
address to trigger the attack. \circled{2}~The Trojan generates a malicious 
GETX packet for the target address. \circled{3}~The directory receives the GETX 
request, broadcasts an invalidation to all cores, and waits for all cores to 
send acknowledgments (\texttt{ACK}s). \circled{4}~The directory forwards the 
data and all \texttt{ACK}s to the compromised core. \circled{5}~The Trojan 
blocks the local directory from seeing any response from the directory or 
cores, waiting to receive all \texttt{ACK}s. \circled{6}~Once all \texttt{ACK}s
are received, the Trojan can access the data, since the directory considers the 
compromised core as the data owner.



\textbf{Phase 2, Writing Malicious Data:}
Once access permissions are acquired, the global directory assumes
that the Trojan's core is the exclusive owner of the data.
%
Figure~\ref{fig:forging_attack}(bottom) illustrates Phase 2 of the
attack.  This phase allows the Trojan to mimic the legal operations
that enable writing to main memory as if the core was evicting the
data after modifying it.  The steps of the attack are as follows:
\circled{1}~Once the Trojan receives the final ACK, the requests to
the target address are unblocked.  \circled{2}~The Trojan immediately
sends a PUTX to the directory to indicate that it is ``evicting''
modified data. \circled{3}~The directory responds with a
\texttt{WRITEBACK\_ACKNOWLEDGEMENT}, allowing the Trojan to proceed
with ``evicting'' the maliciously changed dirty data. \circled{4}~The
Trojan responds to the \texttt{WRITEBACK\_ACKNOWLEDGEMENT} with a
\texttt{WRITEBACK\_EXCLUSIVE\_DIRTY} response containing the malicious
data. \circled{5}~The data is written to memory.

\subsection{Results}

\begin{figure}[ht] 
  \centering
  \subfigure[Data received by the victim when the Trojan is \textbf{not}
    activated. The application reads an alternating sequence of `1' and `0.']
    {\label{fig:notrojan_result}\includegraphics[width=0.48\columnwidth]
    {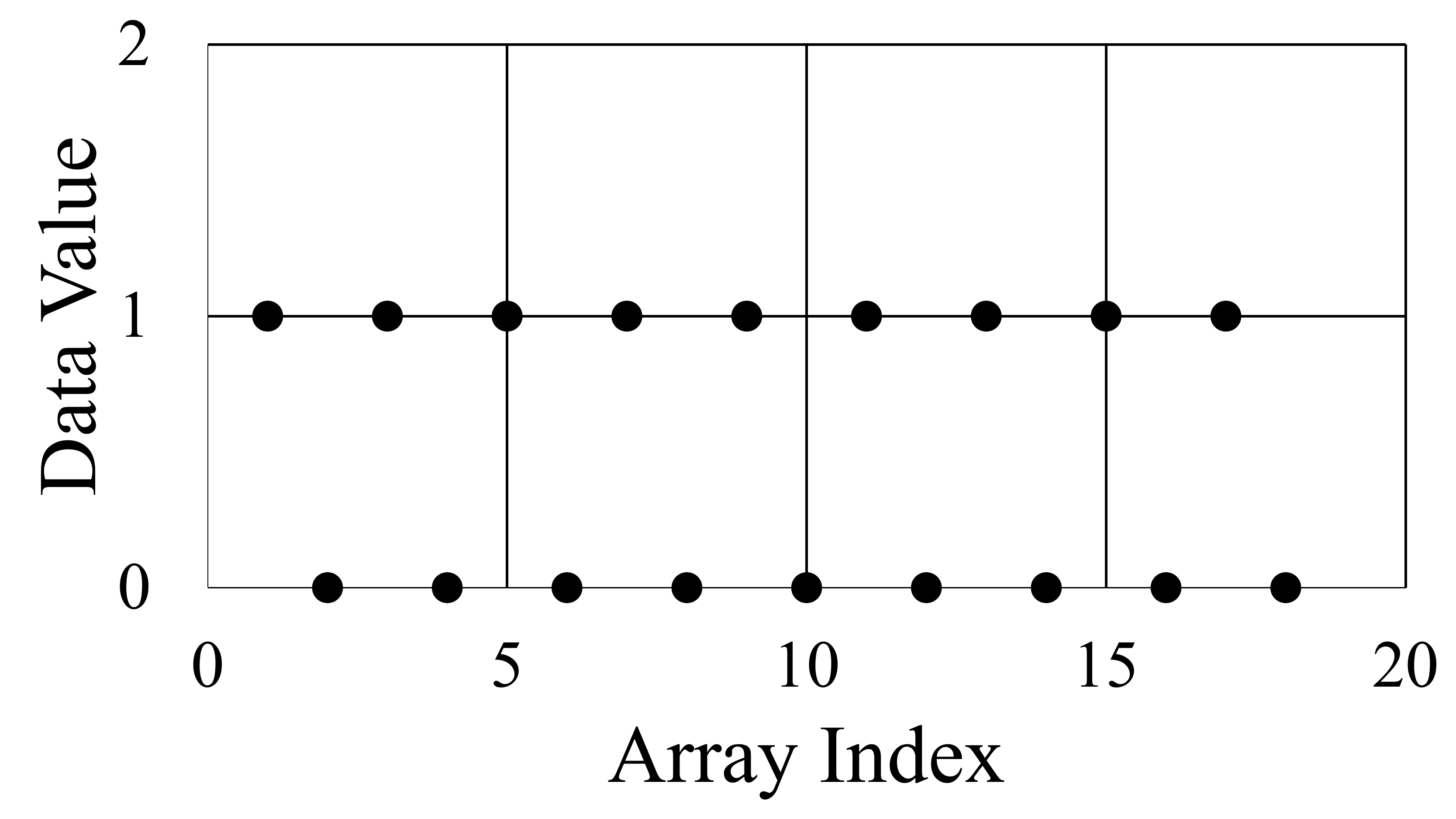}}
  \hfill
  \subfigure[Data received after the Trojan has completed its
    attack. The first entry in the array is now set to `5,' instead of
    the expected `1.']
    {\label{fig:trojan_result}\includegraphics[width=0.48\columnwidth]
    {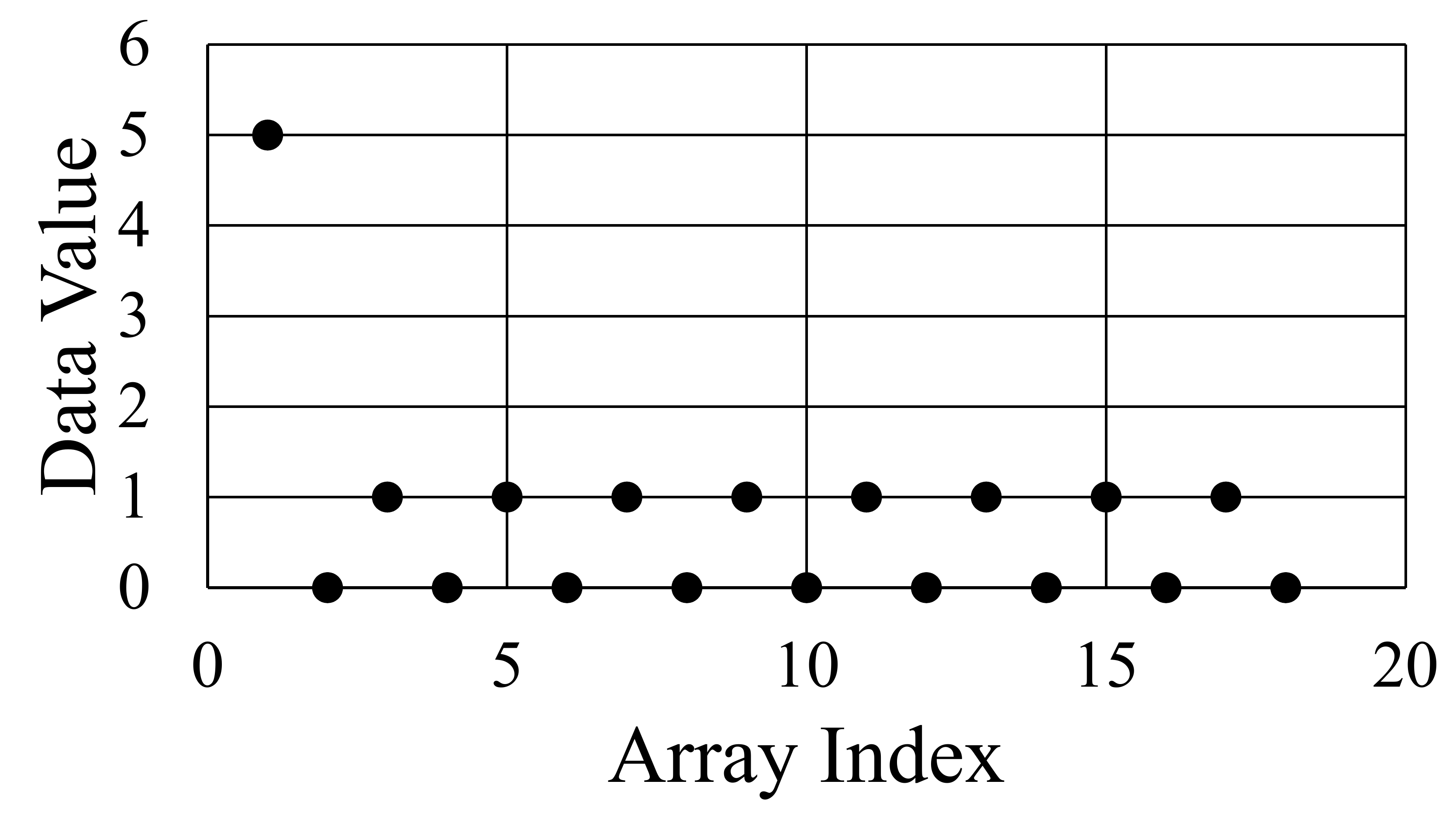}}

  \caption{Data values as seen by the victim.}

\label{fig:attack_results}
\end{figure}

We evaluate the Trojans in \textit{gem5}, targeting a victim which
iterates over an array to set each value to '1' or '0' and then reads
the array to compute a sum. Figure~\ref{fig:notrojan_result} shows the
data the victim process observes without the Trojan enabled. The
victim writes `0' or `1' to various locations in its data array and
then re-reads these locations, seeing the expected data
values. Figure~\ref{fig:trojan_result} shows the data the victim
receives when it attempts to read the data array after writing to all
indexes.  The \emph{Forging Attack} successfully modifies the data
array's first value, which the victim then reads unknowingly of the
manipulation.  This demonstrates our Trojan can manipulate the
coherence system to modify data that another application is operating
on, even without requiring shared memory access.

Unlike prior work, which focuses on Trojans modifying
packets~\cite{ancajas_2014,boraten_2016,khan_2021,prasad_2017}, we
leverage the coherence mechanism itself to modify data in memory that
is \emph{never touched} and \emph{not owned} by the chiplet containing
the Trojan. Our attack does not require the data to be in the compromised 
core's caches. Generating and blocking specific coherence messages allows the 
Trojan to mislead the global directory about the state and ownership of the 
targeted data.

\section{Conclusion and Future Defenses}\label{sec:conclusion}

As industry moves toward chiplet-based designs, hardware Trojans pose a 
significant threat to security. These systems will rely heavily on coherence to 
ensure that data remains up-to-date in all components, making the coherence 
protocol an attractive target. Critically, unlike prior work, which focuses 
only on packet modifications, we show that a coherence-centric Trojan attack 
can modify memory that is not even owned by the compromised chiplet. We provide 
an example of a complex Trojan implementation that modifies memory without 
relying on malicious software components. This work highlights the need for 
mechanisms to protect the coherence scheme from these novel attacks.

Detecting Trojans during chiplet manufacturing is challenging
considering the complexity of individual IPs.  Defenses against
hardware Trojan exploiting a system's coherence mechanisms could
implement runtime monitoring to identify malicious behavior
originating from a particular chiplet.
A benefit of 2.5D integration is that the components are usually sourced from 
vendors and then integrated onto an interposer layer at a separate foundry than 
each IP's manufacturing~\cite{rocketchip}. Requiring an interposer's
manufacturing and integration by a trustworthy facility could allow
the 2.5D interposer to act as a hardware root of trust that can embed
security features. Embedding the security features into the
interposer layer could allow defenders to observe coherence packets
and securely control data flow freely. We plan to explore these
themes in our future work.


\ifCLASSOPTIONcaptionsoff
  \newpage
\fi

\bibliographystyle{ieeetr}
\bibliography{ref}

\begin{thebibliography}{10}

\bibitem{naffziger21}
S.~Naffziger, N.~Beck, T.~Burd, K.~Lepak, G.~H. Loh, M.~Subramony, and
  S.~White, ``Pioneering chiplet technology and design for the {AMD} {EPYC} and
  {Ryzen} processor families : Industrial product,'' in {\em ACM/IEEE ISCA},
  pp.~57--70, 2021.

\bibitem{cxl}
``{Compute Express Link (CXL)}, {www.computeexpresslink.org}.''
\newblock Accessed: 2022-05-18.

\bibitem{bhunia2018hardware}
S.~Bhunia and M.~Tehranipoor, {\em The Hardware Trojan War}.
\newblock Springer, 2018.

\bibitem{9061138}
M.~N.~I. {Khan}, A.~{De}, and S.~{Ghosh}, ``Cache-out: Leaking cache memory
  using hardware {Trojan},'' {\em IEEE TVSLI}, vol.~28, no.~6, pp.~1461--1470,
  2020.

\bibitem{7753274}
M.~{Bidmeshki}, G.~R. {Reddy}, L.~{Zhou}, J.~{Rajendran}, and Y.~{Makris},
  ``Hardware-based attacks to compromise the cryptographic security of an
  election system,'' in {\em IEEE ICCD}, pp.~153--156, 2016.

\bibitem{7926975}
M.~{Kim}, S.~{Kong}, B.~{Hong}, L.~{Xu}, W.~{Shi}, and T.~{Suh}, ``Evaluating
  coherence-exploiting hardware {Trojan},'' in {\em IEEE DATE}, pp.~157--162,
  2017.

\bibitem{basak2017ifs}
A.~Basak, S.~Bhunia, T.~Tkacik, and S.~Ray, ``Security assurance for
  system-on-chip designs with untrusted {IPs},'' {\em IEEE TIFS}, vol.~12,
  no.~7, pp.~1515--1528, 2017.

\bibitem{Conway_2010}
P.~{Conway}, N.~{Kalyanasundharam}, G.~{Donley}, K.~{Lepak}, and B.~{Hughes},
  ``Cache hierarchy and memory subsystem of the {AMD} {Opteron} processor,''
  {\em IEEE Micro}, vol.~30, no.~2, pp.~16--29, 2010.

\bibitem{rocketchip}
J.~Kim {\em et~al.}, ``Architecture, chip, and package co-design flow for
  {2.5D} {IC} design enabling heterogeneous {IP} reuse,'' in {\em ACM/IEEE
  DAC}, pp.~1--6, 2019.

\bibitem{ancajas_2014}
D.~M. Ancajas, K.~Chakraborty, and S.~Roy, ``{Fort-NoCs}: Mitigating the threat
  of a compromised {NoC},'' in {\em ACM/EDAC/IEEE DAC}, pp.~1--6, 2014.

\bibitem{boraten_2016}
T.~Boraten and A.~K. Kodi, ``Mitigation of denial of service attack with
  hardware {Trojans} in {NoC} architectures,'' in {\em IEEE IPDPS},
  pp.~1091--1100, 2016.

\bibitem{khan_2021}
M.~H. Khan, R.~Gupta, J.~Jose, and S.~Nandi, ``Dead flit attack on {NoC} by
  hardware {Trojan} and its impact analysis,'' in {\em ACM NoCArc}, pp.~10--15,
  2021.

\bibitem{prasad_2017}
N.~Prasad, R.~Karmakar, S.~Chattopadhyay, and I.~Chakrabarti, ``Runtime
  mitigation of illegal packet request attacks in networks-on-chip,'' in {\em
  IEEE ISCAS}, pp.~1--4, 2017.

\end{thebibliography}


%




\end{document}